\documentclass[aps,prd,twocolumn,groupedaddress,showpacs,amsmath,amssymb]{revtex4}
\usepackage{graphicx,epsf}
\usepackage[]{latexsym}
\newcommand{\be}{\begin{eqnarray}}
\newcommand{\ee}{\end{eqnarray}}

\newcommand{\ket}[1]{\mbox{$\mid #1\,\rangle$}}

\begin{document}
\title{Trans-Planckian footprints in inflationary cosmology}
\author{G.L.~Alberghi}
\email{alberghi@bo.infn.it}
\author{R.~Casadio}
\email{casadio@bo.infn.it}
\author{A.~Tronconi}
\email{tronconi@bo.infn.it}
\affiliation{Dipartimento di Fisica, Universit\`a di Bologna, and
I.N.F.N, Sezione di Bologna, Via Irnerio 46, 40126 Bologna, Italy.}
\begin{abstract}
We consider a minimum uncertainty vacuum choice at a fixed energy
scale $\Lambda$ as an effective description of trans-Planckian
physics, and discuss its implications for the linear perturbations
of a massless scalar field in power-law inflationary models.
We find possible effects with a magnitude of order $H/\Lambda$
in the power spectrum, in analogy with previous results for de-Sitter
space-time.
\end{abstract}
\maketitle
\section{Introduction}
\label{intro}
Inflation has nowadays become a standard ingredient for the
description of the early Universe (see, e.g., Refs.~\cite{infla}).
In fact, it solves some of the problems of the standard big-bang
scenario and also makes predictions about cosmic microwave
background radiation (CMBR) anisotropies which are being measured
with higher and higher precision.
Further, it has been recently suggested that inflation might
provide a window towards {\em trans\/}-Planckian physics
\cite{branden} (for a partial list of subsequent works on this
subject, see Refs.~\cite{tp,kempf,kleban,easther,b-m,ulf}).
The reason for this is that inflation magnifies all quantum
fluctuations and, therefore, red-shifts originally trans-Planckian
frequencies down to the range of low energy physics.
This causes two main concerns:
first of all, there is currently no universally accepted
(if at all) theory of quantum gravity which allows us to describe
the original quantum fluctuations in such an high energy regime;
further, it is not clear whether the red-shifted trans-Planckian
frequencies can indeed be observed with the precision of present
and future experiments.
\par
Regarding the first problem, one can take the pragmatic approach
of modern renormalization theory and assume that quantum
fluctuations are effectively described by quantum field theory
after they have been red-shifted below the scale of quantum
gravity, henceforth called $\Lambda$, and forget about their
previous dynamics.
Further, one can also take $\Lambda$ as a constant throughout
the evolution of the (homogeneous and isotropic) Universe, thus
implicitly assuming the existence of some preferred reference
frame (class of ``cosmological'' observers).
The second problem is instead more of a phenomenological interest
and needs actual investigation to find the size of corrections
to the CMBR.
It then seems that the answer depends on the details of the
model that one considers and no general consensus has been reached
so far.
In fact, in Refs.~\cite{kempf,kleban} it is claimed that such
corrections can be at most of order $(H/\Lambda)^2$, where $H$
is the Hubble parameter, hence too small to be detected.
However, corrections are estimated of order $H/\Lambda$ in
Refs.~\cite{easther,b-m,ulf}.
Let us note that the first problem also plays an important role
in this phenomenological respect, since it is the unknown
trans-Planckian physics which fixes the ``initial conditions''
for the effective field theory description.
\par
In Ref.~\cite{ulf}, a principle of least uncertainty on the
quantum fluctuations at the time of emergence from the Planckian
domain (when the physical momentum $p\sim\Lambda$) was imposed.
Without a good understanding of physics at the Planck scale,
this can be regarded as an empirical way of accounting for
new physics.
Such a prescription fixes the initial vacuum (independently)
for all frequency modes, and subsequent evolution is then obtained
in the {\em sub\/}-Planckian domain by means of standard Bogolubov
transformations (of course, neglecting the back-reaction)
in de-Sitter space-time.
In the present paper, we apply the same approach as in
Ref.~\cite{ulf} to power-law inflation.
This will allow us to check the final result against an
inflationary model with time-dependent Hubble parameter.
\section{Sub-Planckian effective theory}
\setcounter{equation}{0}
\label{general}
On the homogeneous and isotropic background
\be
ds^2=a^2(\eta)\,\left[-d\eta^2+dx^2+dy^2+dz^2\right]
\ ,
\ee
the spatial Fourier components of the (rescaled) scalar field
$\mu=a\,\phi$ (as well as tensor perturbations $\mu_T$) satisfy
\be
\mu''_k+\left(k^2-{a''\over a}\right)\,\mu_k=0
\ ,
\label{kg}
\ee
where primes denote derivative with respect to the conformal
time $-\infty<\eta<0$.
\par
The index $k$ is related to the physical momentum $p$ by $k=a\,p$.
Thus, a given mode with energy above the Planck scale in the far
past would cross the fundamental scale $\Lambda$ at the time
$\eta_k$ when
\be
k=a(\eta_k)\,\Lambda
\label{re}
\ .
\ee
Strictly speaking, it is incorrect to regard such a mode as existing
for $\eta<\eta_k$, since we do not have a theory for that case.
What we will in fact consider is just the evolution for $\eta>\eta_k$.
\subsection{Minimum uncertainty principle}
Following Ref.~\cite{ulf}, we shall impose that the mode $k$ is put into
being with minimum uncertainty at $\eta=\eta_k$, that is the vacuum
satisfies in the Heisenberg picture (for the details see, e.g.,
Ref.~\cite{polarski})
\be
\hat\pi_k(\eta_k)\ket{0}=i\,k\,\hat\mu_k(\eta_k)\ket{0}
\ ,
\label{min}
\ee
where 
\be
\pi_k=\mu_k'-{a'\over a}\,\mu_k
\ee
is the Fourier component of the momentum $\pi$ conjugate to $\mu$.
We can write the scalar field and momentum at all times in terms of
annihilation and creation operators for time dependent oscillators
\be
\begin{array}{l}
\hat \mu _k (\eta) = { 1 \over \sqrt{2k}}\,
 \left[ \hat a_k (\eta) + \hat a^{\dagger}_{-k} (\eta) \right]
\\
\\
\hat \pi _k (\eta) = -i \sqrt{ {k \over 2}}\, 
 \left[ \hat a_k (\eta) - \hat a^{\dagger}_{-k} (\eta) \right]
\ .
\end{array}
\label{definitions}
\ee
The oscillators can be expressed in terms of their values at the
time $ \eta _k $ through a Bogoliubov transformation
\be
\begin{array}{l}
\hat a_k(\eta)=u_k(\eta)\,\hat a_k(\eta_k)
+v_k(\eta)\,\hat a^\dagger_{-k}(\eta_k)
\\
\\
\hat a_{-k}^\dagger(\eta)=u_k^*(\eta)\,\hat a_{-k}^\dagger(\eta_k)
+v_k^*(\eta)\,\hat a_{k}(\eta_k)
\ .
\end{array}
\ee
Substituting this expression in (\ref{definitions}) we obtain
\be
\begin{array}{l}
\hat\mu_k(\eta)=f_k(\eta)\,\hat a_k(\eta_k)
+f_k^*(\eta)\,\hat a^\dagger_{-k}(\eta_k)
\\
\\
i\,\hat \pi_k(\eta)=g_k(\eta)\,\hat a_k(\eta_k)
-g_k^*(\eta)\,\hat a^\dagger_{-k}(\eta_k)
\ .
\end{array}
\ee
where
\be
\begin{array}{l}
f_k ( \eta) = { 1 \over \sqrt{2k}}\,
\left[ u_k( \eta) + v^{*} _k (\eta)\right]
\\
\\
g_k (\eta) = \sqrt{ {k \over 2} }\,
\left[ u_k(\eta) - v_k^{*} (\eta) \right]
\ ,
\end{array}
\ee
and $ f_k (\eta) $ is a solution of the mode equation (\ref{kg}).
The condition (\ref{min}) then reads
\be
v_k(\eta_k)=\sqrt{k\over 2}\,f_k^*(\eta_k)
-{1\over \sqrt{2\,k}}\,g_k^*(\eta_k)
=0
\ .
\label{ini}
\ee
\par
This requirement, together with the normalization condition
\be
|u_k|^2-|v_k|^2=1
\ ,
\label{norm}
\ee
is sufficient to determine uniquely the initial state at
$\eta=\eta_k$.
The subsequent time evolution is then straightforward and one
can estimate the power spectrum of fluctuations at a later
time $\eta\gg\eta_k$ after the end of inflation,
\be
P_\phi={P_\mu\over a^2}
={k^3\over 2\,\pi^2\,a^2}\,\left|f_k(\eta)\right|^2
\ .
\ee
\par
The above general formalism was applied to de-Sitter space-time
in Ref.~\cite{ulf}.
For that case, one has $a=-1/H\,\eta$ and the nice feature
follows that
\be
k\,\eta_k=-{\Lambda\over H}
\ee
is a constant independent of $k$.
This, in turn, allows to obtain an analytic expression for the
initial state which satisfies Eq.~(\ref{ini}) by suitably expanding
for $H/\Lambda$ small (i.e., $\eta_k\to-\infty$ for all $k$).
We shall instead consider power-law inflation, where such a 
simplification does not occur.
\subsection{Power-law inflation}
In the proper time $dt=a\,d\eta$, power-law inflation is given by
a scale factor $a\sim t^p$, in which $t_{\rm p}<t<t_{\rm o}$, with
$t_{\rm p}$ of the order of the Planck time,
$t_{\rm o}\gg t_{\rm p}$ is the time of the end of
inflation, and $p\gg 1$ \cite{easther}.
Upon changing to the conformal time, one obtains for the scale factor
\be
a(\eta)=\left(\bar\eta\over\eta\right)^q
\ ,
\label{a}
\ee
where $q=p/(p-1)$, $\eta_{\rm p}<\eta\le\eta_{\rm o}<0$
($\eta_{\rm o}$ is the end of inflation) and the Hubble parameter
is given by
\be
H(\eta)=-q\,{\eta^{q-1}\over\bar\eta^q}
\ .
\ee
The condition (\ref{re}) now becomes
\be
k\,\eta_k=\bar\eta\,\Lambda^{1\over q}\,k^{1-{1\over q}}
\ .
\label{etak}
\ee
Since the right hand side depends on $k$ (unless $q=1$), it can
be large or small depending on $k$, and an expansion for $-k\,\eta_k$
large is not generally valid.
\par
For the scale factor (\ref{a}) one has
\be
{a''\over a}={q\,(q+1)\over\eta^2}
\ ,
\ee
and Eq.~(\ref{kg}) can be solved exactly.
One can write the general solution as
\be
f_k=A_k\,\sqrt{-\eta}\,J_{q+{1\over 2}}(-k\,\eta)
+B_k\,\sqrt{-\eta}\,Y_{q+{1\over 2}}(-k\,\eta)
\ ,
\ee
where $J_\nu$ and $Y_\nu$ are Bessel functions of the first and
second kind~\footnote{We remark that such functions are real in
the chosen domain of $\eta$.}, and $A_k$ and $B_k$ are complex
constants.
The Bogolubov coefficients are then given by
\be
\!\!\!\!\!
u_k\!\!\!&=&\!\!\!
\sqrt{-{k\,\eta\over 2}}\,\left[
A_k\,J_{q+{1\over 2}}(-k\,\eta)+B_k\,Y_{q+{1\over 2}}(-k\,\eta)
\right.
\nonumber
\\
&&
\left.
-i\left(A_k\,J_{q-{1\over 2}}(-k\,\eta)
+B_k\,J_{q-{1\over 2}}(-k\,\eta)\right)\right]
\nonumber
\\
&&
\\
\!\!\!\!\!
v_k^*\!\!\!&=&\!\!\!
\sqrt{-{k\,\eta\over 2}}\,\left[
A_k\,J_{q+{1\over 2}}(-k\,\eta)+B_k\,Y_{q+{1\over 2}}(-k\,\eta)
\right.
\nonumber
\\
&&
\left.
+i\left(A_k\,J_{q-{1\over 2}}(-k\,\eta)
+B_k\,Y_{q-{1\over 2}}(-k\,\eta)\right)\right]
\ .
\nonumber
\ee
\par
The constants $A_k$ and $B_k$ can now be fixed by imposing the
normalization condition (\ref{norm}) and Eq.~(\ref{ini}).
From Eq.~(\ref{norm}) one obtains
\be
A_k\,B_k^*-A_k^*\,B_k=-i\,{\pi\over 2}
\ ,
\ee
and from Eq.~(\ref{ini}),
\be
A_k=-{\bar Y_{q+{1\over 2}}+i\,\bar Y_{q-{1\over 2}}\over
\bar J_{q+{1\over 2}}+i\,\bar J_{q-{1\over 2}}}\,B_k
\ ,
\ee
where $\bar J_\nu\equiv J_\nu(-k\,\eta_k)$ and
$\bar Y_\nu\equiv Y_\nu(-k\,\eta_k)$.
From the combined equations one then obtains
\be
&&|A_k|^2=-{\pi^2\over 8}\,k\,\eta_k\,\left[\bar Y_{q+{1\over 2}}^2
+\bar Y_{q-{1\over 2}}^2\right]
\nonumber
\\
&&
|B_k|^2=-{\pi^2\over 8}\,k\,\eta_k\,\left[\bar J_{q+{1\over 2}}^2
+\bar J_{q-{1\over 2}}^2\right]
\\
&&
{\rm Re}\left(A_k\,B_k^*\right)=
{\pi^2\over 8}\,k\,\eta_k\,\left(
\bar Y_{q+{1\over 2}}\,\bar J_{q+{1\over 2}}
+\bar Y_{q-{1\over 2}}\,\bar J_{q-{1\over 2}}\right)
\ .
\nonumber
\ee
We are finally in the position to compute the exact power spectrum
at the time $\eta\le\eta_{\rm o}$, which is given by
\begin{widetext}
\be
P_\phi\!\!\!&=&\!\!\!
{\eta_k\,\eta^{2\,q+1}\,k^4\over 16\,\bar\eta^{2\,q}}
\left\{\left[\bar Y_{q+{1\over 2}}\,J_{q+{1\over 2}}(-k\,\eta)
-\bar J_{q+{1\over 2}}\,Y_{q+{1\over 2}}(-k\,\eta)\right]^2
+\left[\bar Y_{q-{1\over 2}}\,J_{q+{1\over 2}}(-k\,\eta)
-\bar J_{q-{1\over 2}}\,Y_{q+{1\over 2}}(-k\,\eta)\right]^2
\right\}
\ .
\label{Pexact}
\ee
\end{widetext}
The above expression can then be estimated for $\eta=\eta_{\rm o}$
(end of inflation) and $\eta_{\rm o}\to 0^-$.
Since for $-k\,\eta_{\rm o}\ll 1$, the Bessel $Y_{q+{1\over 2}}$
dominates, one obtains, to leading order,
\be
P_\phi\simeq 
{k^{3-2\,q}\,|\eta_k|\over 2^{3-2\,q}\,|\bar\eta|^{2\,q}}
{\bar J_{q+{1\over 2}}^2+\bar J_{q-{1\over 2}}^2\over
\sin^2\left(\pi\left(q+{1\over 2}\right)\right)\,
\Gamma^2\left({1\over 2}-q\right)}
\ .
\ee
If one further takes the limit $k\,\eta_k\to-\infty$ and
expands to leading order for $k$ small, the power spectrum
becomes
\be
P_\phi&\simeq& {2^{2\,q-2}\,k^{2-2\,q}
\over\pi\,|\bar\eta|^{2\,q}\,\cos^2(\pi\,q)\,
\Gamma^2\left({1\over 2}-q\right)}
\nonumber
\\
&&\times
\left[1-{H_k\over \Lambda}\,
\sin(2\,\bar\eta\,\Lambda^{1\over q}\,k^{1-{1\over q}}
+\pi\,q)\right]
\nonumber
\\
&=&P_{\rm PL}\,\left[1- {H_k \over \Lambda} 
  \sin \left(q\, {2\,\Lambda \over H_k} + q\,\pi \right) \right]
\ ,
\label{Papp}
\ee
where $H_k\equiv H(\eta_k)$ and we have factored out the expression
$P_{\rm PL}\sim k^{2-2\,q}$ of the spectrum for power-law
inflation~\cite{abbott} in the small $k\,\eta_{\rm o}$ regime
(super-horizon scales) \cite{martin}.
This result is thus in agreement with what was obtained for de-Sitter
space-time in Ref.~\cite{ulf}, as one can easily see by taking the limit 
$ q \to 1 $ ($p\to\infty$).
\par
\begin{figure}[t]
\epsfxsize=3.2in
\epsfbox{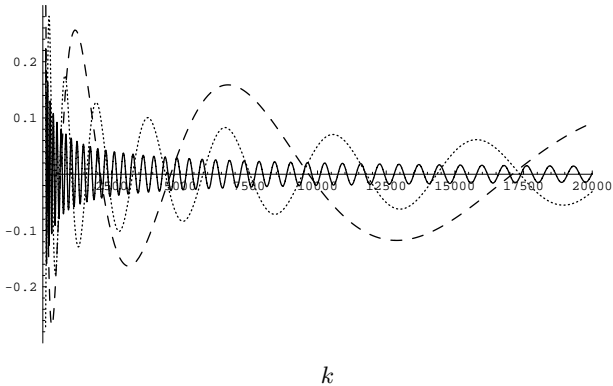}
\raisebox{2.0cm}{$k$}
\caption{The ratio $R_q$ for $P_\phi$ in Eq.~(\ref{Pexact}) and $q=2$
(solid line), $q=3/2$ (dotted line) and $q=4/3$ (dashed line).
The momentum index $k$ is in units with $\Lambda=\bar\eta=1$ and the
regions of physical interest are those for $k\gtrsim 10^2$ ($q=2$),
$k\gtrsim 10^3$ ($q=3/2$) and $k\gtrsim 10^4$ ($q=4/3$).
\label{spectra}}
\end{figure}

However, as we mentioned previously, $k\,\eta_k$ is not
independent of $k$ [see Eq.~(\ref{etak})].
The above expression therefore does not hold for all $k$, but just
for those such that $-k\,\eta_k$ is large.
Since it is very difficult to obtain general analytic estimates of
the exact power spectrum for general values of $k$, in
Fig.~\ref{spectra} we plot, for the exact expression of
$P_\phi$ in Eq.~(\ref{Pexact}), the ratio
\be
R_q={P_\phi-P_{\rm PL}\over P_{\rm PL}}
\ ,
\ee
for $q=2$, $3/2$ and $4/3$ (similar results are obtained for all
values of $q\not=1$).
It is clear that for small $k$ the oscillations in $P_\phi$ are
relatively large around $P_{\rm PL}$, and this is precisely due to
the dependence of $k\,\eta_k$ on $k$.
The oscillations are then progressively damped for large $k$
according to the approximate expression in Eq.~(\ref{Papp})
(and analogously to what is found in de-Sitter \cite{ulf}).
Note also that for increasing $p$ (i.e.~$q\to 1^+$), the wavelength
of oscillations increases, as is shown in the approximation
(\ref{Papp}).
Of course, one must keep in mind that only sub-horizon scales
matter at the time $\eta_k$, for which $k\gg a\,H$, that is
$|k\,\eta_k|\gg q$ (say of order $\lambda$).
Hence, the relevant regions for different values of $q$ are those
with $k\gtrsim \lambda^{q/(q-1)}$.
In Fig.~\ref{spectra} we have set $\lambda=10$ in order to obtain
reasonably overlapping ranges, and the amplitude of the oscillations
turns out to be of the order of a few percents inside the physical
ranges (larger values of $\lambda$ imply smaller oscillations). 
\section{Conclusions}
\setcounter{equation}{0}
\label{conc}
We considered a minimum uncertainty principle to fix, at an energy
scale $\Lambda$, the vacuum of an effective (low energy) field theory.
Such prescription involves the cut-off scale $\Lambda$ for dealing
with trans-Planckian energies, which therefore enters into the
power spectrum of perturbations at later times.
We have shown in some details that a $\Lambda$ of the order of the
Planck scale can affect appreciably the spectrum [see Eq.~(\ref{Papp})
and Fig.~\ref{spectra}], in agreement with Refs.~\cite{easther,b-m,ulf}
by introducing a modulation of the spectrum, as may be clearly seen
from the figure.
This is a clear indication that trans-Planckian physics can lead to
observable predictions in the cosmological models.
We feel this is further evidence for the fact that trans-Planckian
physics cannot be safely ignored in determining observable quantities
such as features of the CMBR.
\begin{acknowledgments}
We would like to thank R.~Brandenberger, R.~Brout, F.~Finelli,
F.~di~Marco and G.~Venturi for useful discussions.
G.L.~A.~would also like to thank D.A.~Lowe for interesting discussions
on the subject and the High~Energy~Theory~Group of Brown University
for hospitality during the early stages of the present work.
R.~C.~would also like to thank L.~Mersini for many discussions on
the subject.
\end{acknowledgments}
\end{document}